\documentclass[12pt,fleqn]{article}
\input{epsf}
\usepackage{amsmath}
\usepackage{latexsym}
\usepackage{amssymb}
\newlength{\dinwidth}
\newlength{\dinmargin}
\newcommand{\resection}[1]{
\renewcommand{\theequation}{\thesection.\arabic{equation}}
\setcounter{equation}{0}\section{#1}}
\newcommand{\resub}[1]{
\setcounter{equation}{0}
\renewcommand{\theequation}{\thesubsection.\arabic{equation}}
\subsection{#1}}

\newcommand{\appsectio}[1]{\setcounter{section}{0}
         \addtocounter{section}{1} \setcounter{equation}{0}
                         \section*{Appendice \Alph{section}}}
\newcommand{\appsection}[1]{\addtocounter{section}{1} \setcounter{equation}{0}
                         \section*{Appendice \Alph{section}}}
\renewcommand{\theequation}{\thesection.\arabic{equation}}

\newcommand{\f}[2]{\frac{#1}{#2}}

\def\p{\phi}
\def\d{\partial}

\def\xb{\bar{x}}
\def\kh{\hat{k}}
\def\ph{\hat{p}}
\def\vt{\tilde{v}}
\def\Ct{\tilde{C}}
\def\vb{\bar{v}}
\def\Cb{\bar{C}}
\def\pts{(\phi^3)_6}
\def\dk{\frac{d^6k}{(2\pi)^6}}
\def\dkd{\frac{d^D k}{(2\pi)^D}}
\def\G{\Gamma}
\def\D{\Delta}
\def\g{\gamma}
\def\eps{\epsilon}
\def\tge{t^\g_\eps}
\def\UU{\cal U}
\def\UUT{\tilde {\cal U}}
\def\Ut{{\cal U}_t}
\def\Un{{\cal U}^n}
\def\hs{\hspace{.5mm}}
\newcommand{\nn}{\nonumber}
\renewcommand{\thefootnote}{\dag}
\begin{document}
\vspace*{1.5cm}
\begin{center}

{\huge Cut Vertices and Semi-Inclusive\\}
\vspace*{.5truecm} 
{\huge Deep Inelastic Processes}

\vskip 1truecm

{\large M. Grazzini}

\vspace*{.1truecm}
{\em Dipartimento di Fisica, Universit\`a di Parma and\\ INFN, Gruppo Collegato di Parma, I-43100 Parma, Italy\\}
\vskip 1truecm
October 4, 1997
\end{center}

\renewcommand{\thefootnote}{\arabic{footnote}}
\setcounter{footnote}{0}
\vskip .5cm

 \begin{abstract}
 Cut vertices, a generalization of matrix elements of local
 operators, are revisited, and an expansion in
 terms of minimally subtracted
 cut vertices is formulated.
 An extension of the formalism to deal with
 semi-inclusive deep inelastic processes in the target
 fragmentation region is
 explicitly constructed. 
 The problem of factorization is discussed in detail. 
 \end{abstract}

\vskip 2.5truecm

\begin{center}
UPRF-97-016\\
\end{center}

\newpage

\setcounter{page}{1}

\resection{Introduction}

Factorization theorems \cite{fact,css} are an essential tool in
QCD hard processes
since they allow to separate
short-distance from long-distance effects.
The former, due to asymptotic freedom, are calculable in perturbation
theory, while the latter can be parameterized into phenomenological
distributions.

Factorization theorems are proved by showing that cross sections
develop
collinear singularities occurring in a universal factorized fashion,
while
soft
singularities,
which could in principle spoil such a picture, cancel out.

Operator Product Expansion (OPE) \cite{wilson,zim} as well is a powerful method to separate
short-distance from long-distance effects but it
has unfortunately found practical applications only in the context of
inclusive
Deep Inelastic Scattering (DIS).
This stems from
the fact that OPE works for amplitudes in particular
kinematic
regions whereas to describe
physical processes one needs a prediction for
cross sections, i.e. cut amplitudes.
However there is a generalization of OPE, the {\em cut vertex expansion},
originally proposed by Mueller
\cite{mueller}, that allows
to treat processes different from DIS,
since it directly deals with cross sections.
With this method the cross section for DIS ($e^+ e^-\to h+X$)
factorizes
into a spacelike (timelike) cut vertex convoluted with a coefficient function.
The former contains the long-distance behavior while the
latter is
calculable as usual in perturbation theory.
In the context of DIS this expansion is equivalent to the one obtained
by applying OPE, as it can be proved that a spacelike cut vertex
is actually equivalent to a matrix element
of a minimal twist local operator \cite{bau,mun}.
Nevertheless the cut vertex expansion holds also in
$e^+ e^-\to h+X$ and in this case the cut vertex corresponds to
a non-local operator.
In Ref.\cite{bau} a parton model interpretation of
cut vertices was given. In such a picture a spacelike cut vertex is in some
sense equivalent to a parton distribution, whereas a timelike cut vertex
corresponds to a fragmentation function.
The cut vertex technique has been
applied in Ref.\cite{gm} to a
variety of hard processes.

We
remarked that a cut vertex is a (generally non-local) operator
and therefore it needs renormalization. Cut vertex
renormalization was originally proposed in the collinear
subtraction scheme \cite{mueller} but one
can work in the MS scheme
with minor modifications \cite{gupta}
and
here
the coefficient function is
expected to be
analytical in the
zero mass limit \cite{lev}.

Let us now consider
the
semi-inclusive process
$p+J(q)\to p^\prime +X$. In the region in which the transverse
momentum of the observed hadron, or equivalently the momentum transfer
$t=-(p-p^\prime)^2$, is of order of the hard scale $Q^2$,
the cross section
can be written either as a convolution of a
parton density with a hard cross section and a fragmentation function,
or, in the language of cut vertices, as a convolution of a spacelike
with a timelike cut vertex through a coefficient function.
However a similar picture does not work in the limit $t\ll Q^2$
since in that region the cross section is dominated by the target
fragmentation mechanism. 

In a recent paper \cite{cut} an extension of the
cut vertex approach
has been proposed in order to deal with semi-inclusive DIS
in the target fragmentation region.
It has been argued that a generalized cut vertex expansion
holds in the limit $t\ll Q^2$.
This fact has relevant phenomenological consequences.
As shown in Ref.\cite{cut}, once such an expansion is proved,
it is possible to define a new object,
an extended fracture function
${\cal M}_{A,A^\prime}^i(x,z,t,Q^2)$,
just in terms of a new cut vertex. Whereas ordinary
fracture functions \cite{tv} give the probability of
extracting a parton $i$
with momentum fraction $x$ from an incoming hadron $A$,
observing another hadron $A^\prime$
with momentum fraction $z$
in the final state,
extended fracture functions depend on the further scale $t$
at which $A^\prime$ is detected.

In the present paper we want to construct explicitly such a
generalized cut vertex expansion and we will eventually
show that this
program can be carried out with no substantial complication.
The model we will work in is $\pts$, which,
despite its simpler structure, shares several important aspects
with QCD: it is asymptotically free and the
topology of the leading diagrams resembles that of QCD
in a physical gauge.
The cut vertex method, which, as a generalization of OPE,
is based on Zimmermann formalism \cite{zim}, combined with
infrared power counting techniques \cite{css,sterman}, will
allow us to formulate an expansion of the semi-inclusive cross
section in the target fragmentation region into a new
MS cut vertex convoluted with the same coefficient function
appearing in inclusive DIS.

The paper is organized as follows. In Sect. 2 we will review
the cut vertex formalism, first with some examples
and then showing how an expansion in terms of
minimally subtracted cut vertices can be given. In Sect. 3 we
will generalize the cut vertex formalism to semi-inclusive DIS.
In Sect. 4 we will discuss factorization and in Sect. 5 we will
sketch our conclusions.

\resection{A cut vertex expansion}
\resub{Examples}

\begin{figure}[htb]
\begin{center}
\begin{tabular}{c}
\epsfxsize=5truecm
\epsffile{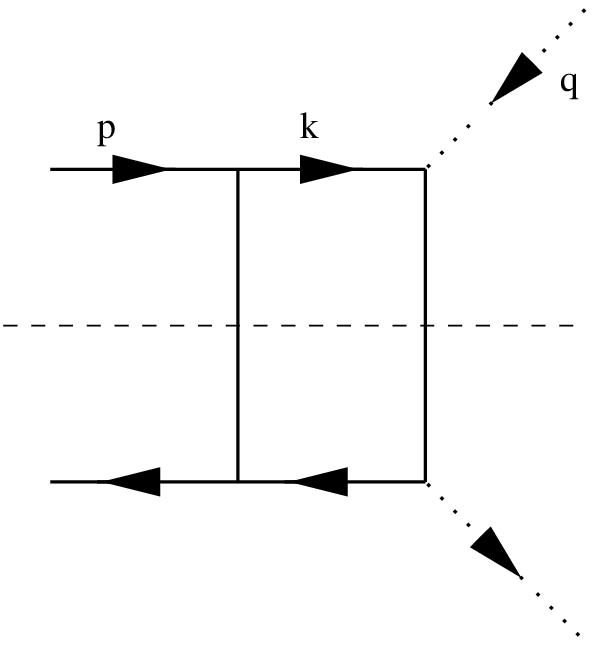}\\
\end{tabular}
\end{center}
\caption{\label{lad}{\em \small A one loop contribution to DIS in $\pts$}}
\end{figure}

Cut vertices
arise
quite naturally when
dealing
with processes
characterized by
a large momentum transfer.
Let us consider DIS in $\phi^3_6$
\begin{equation}
p+J(q)\rightarrow X\nn
\end{equation}
with the current $J=\f{1}{2} \phi^2$.
The structure function is defined as
\begin{equation}
F(p,q)=\f{Q^2}{2\pi} \int e^{iqx} <p|J(x)J(0)|p>.
\end{equation}
We choose a frame in which $p=(p_+,{\bf 0},p_-)$, with $p_+\gg p_-$
and $pq\simeq p_+q_-$.
Given a vector $k=(k_+,{\bf k},k_-)$ define ${\hat k}=(k_+,{\bf0},0)$.
We now consider the ladder contribution to the structure function,
represented in Fig.\ref{lad}.

We have
\begin{equation}
\label{w12}
W_1(p,q)=\int V_1(p,k) M_1(k,q) \dk
\end{equation}
where
\begin{equation}
V_1(p,k)=2\pi \lambda^2\delta_+ ((p-k)^2)\f{1}{k^4}
\end{equation}
\begin{equation}
\label{m1}
M_1(k,q)=Q^2 \delta_+((k+q)^2).
\end{equation}
In the limit $p^2\to 0$ this graph develops a collinear singularity
which corresponds to the configuration in which $k$ is parallel to
$p$.
Hence
in
the integration in (\ref{w12})
two distinct
regions give contribution
in the large $Q^2$ limit\footnote{See for example the
discussion in Ref.\cite{css}}
\begin{align}
& (a)~~~\mbox{collinear:}~~~k^2\ll Q^2 \mbox{ and } k\parallel p\nn\\
& (b)~~~\mbox{UV:}~~~~~~~~~~k^2\sim Q^2.
\end{align}

Factorization is proved by arranging the integral (\ref{w12})
so as
to isolate the contributions in
these regions:
\begin{align}
\label{arranged}
W_1(p,q)&=\int V_1(p,k)M_1(\kh,q)\dk
+\int V_1(\ph,k)\left(M_1(k,q)-M_1(\kh,q)\right)\dk\nn\\
&+\int \Big(V_1(p,k)-V_1(\ph,k)\Big)
\Big(M_1(k,q)-M_1(\kh,q)\Big)\dk.
\end{align}
The first term
dominates
in the collinear region, while
the second is important in the UV region. The third term is
the remainder and is
manifestly
suppressed by a power of $1/Q^2$.
The operation we
performed is called {\em oversubtraction}: it is
a subtraction in addition to standard UV renormalization and
aims at isolating the leading part of the cross
section.

\begin{figure}[htb]
\begin{center}
\begin{tabular}{c}
\epsfxsize=12truecm
\epsffile{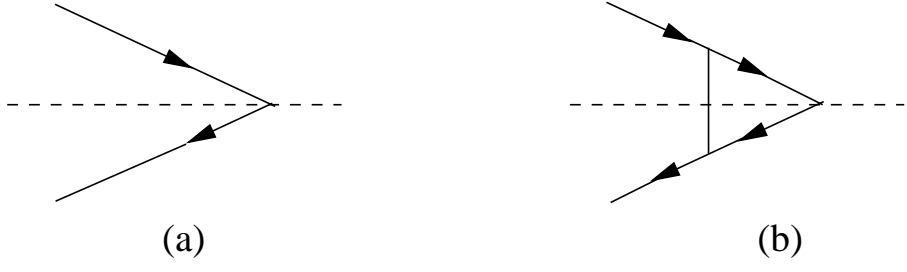}\\
\end{tabular}
\end{center}
\caption{\label{cv}{\em \small (a) The bare cut vertex (b) One of its radiative corrections}}
\end{figure}

It is now straightforward to show that the
first two terms in
(\ref{arranged}) can be written in a factorized form.
If we define the bare cut vertex $v^{(0)}(p^2,x)=\delta(1-x)$ and its
one loop radiative correction (see Fig.\ref{cv})

\begin{figure}[htb]
\begin{center}
\begin{tabular}{c}
\epsfxsize=12truecm
\epsffile{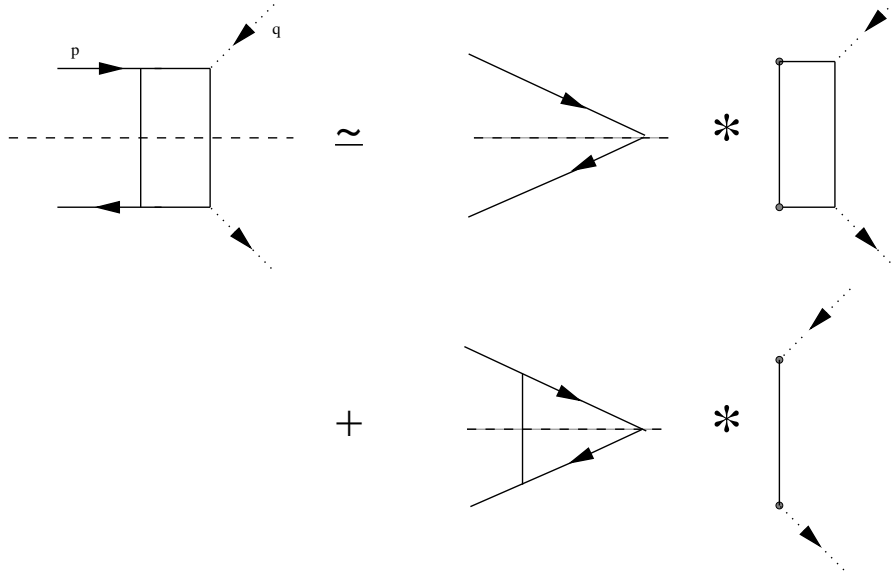}\\
\end{tabular}
\end{center}
\caption{\label{ldmod}{\em \small Factorization of the ladder diagram in $\pts$}}
\end{figure}

\begin{equation}
v^{(1)}(p^2,x)=
\int V_1(p,k) x \delta\left(x-\f{k_+}{p_+}\right)\dk
\end{equation}
we can write
\begin{align}
\label{w1}
W_1(p,q)&\simeq\int \f{du}{u} v^{(1)}(p^2,u)C^{(0)}(x/u,Q^2)
+\int \f{du}{u} v^{(0)}(p^2,u)C^{(1)}_1(x/u,Q^2)\nn\\
&=\int \f{du}{u} v(p^2,u)C_1(x/u,Q^2)\Big|_{order \lambda^2}
\end{align}
where
\begin{equation}
C^{(0)}(x,Q^2)=M_1(0,x,Q^2)
\end{equation}
\begin{equation}
C^{(1)}_1(x,Q^2)=
\int V_1(\ph,k)\left(M_1(k,q)-M_1(\kh,q)\right)\dk
\end{equation}
and
it appears
that factorization works as in Fig.\ref{ldmod}.

As usual a simpler factorized expression can be
obtained by taking moments with respect to $x$.
Defining the Mellin transform as
\begin{equation}
\label{Mellin}
f_\sigma=\int^1_0 dx x^{\sigma-1} f(x)
\end{equation}
we get
\begin{equation}
W_1^\sigma(p^2,Q^2)=v_\sigma(p^2) C_1^\sigma(Q^2).
\end{equation}
Actually both the terms in (\ref{w1}) are UV divergent
expressions. Nevertheless,
due to the
finiteness of the
initial integral, divergences can
be removed by adding and subtracting
a suitable UV counterterm, as we will
clarify
in the next section.

\begin{figure}[htb]
\begin{center}
\begin{tabular}{c}
\epsfxsize=5truecm
\epsffile{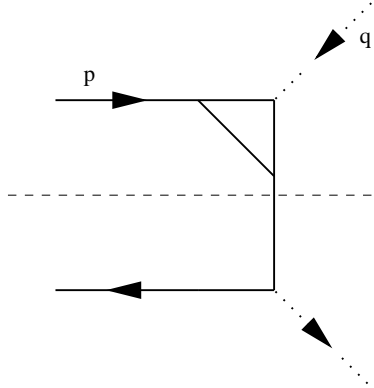}\\
\end{tabular}
\end{center}
\caption{\label{virtual}{\em \small A virtual correction to the DIS in $\pts$}}
\end{figure}

\begin{figure}[htb]
\begin{center}
\begin{tabular}{c}
\epsfxsize=14truecm
\epsffile{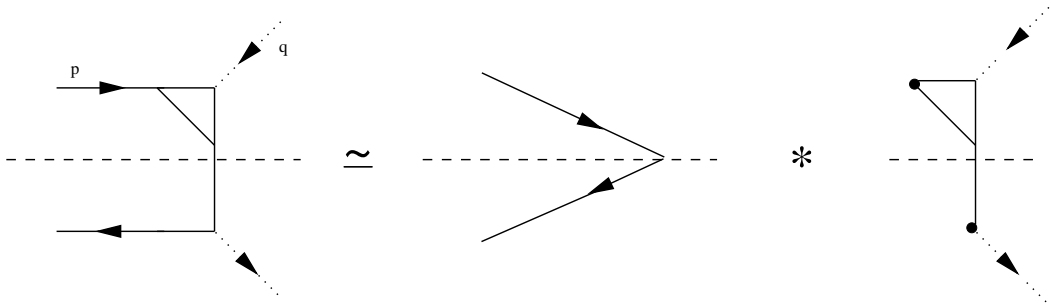}\\
\end{tabular}
\end{center}
\caption{\label{vd}{\em \small Its decomposition in the large $Q^2$ limit}}
\end{figure}

Let us now consider the graph in Fig.\ref{virtual}.
In the large $Q^2$ limit it gives
\begin{equation}
W_2(p,q)=-\f{1}{2}\delta(1-x) \f{\lambda^2}{(4\pi)^3}
\left(\log\f{Q^2}{4\pi\mu^2}+\mbox{finite terms}\right)
\end{equation}
where the proper MS counterterm has been added in order to cure the
ordinary UV divergence.
Whereas in the previous example there was a divergence in the limit
$p^2\to 0$, which is a signal of long-distance dependence, here
the limit $p^2\to 0$ is harmless. 
Hence
\begin{equation}
W_2(p,q)=\int v^{(0)}(p^2,u) C^{(1)}_2(x/u,Q^2)
\end{equation}
with
\begin{equation}
C^{(1)}_2(x,Q^2)=W_2(0,x,Q^2)
\end{equation}
that is factorization works as in Fig.\ref{vd}.
Thus the unique leading contribution is obtained when a large
momentum flows
in the loop, and no long-distance
contribution
arises from this graph which gives only a correction
to the coefficient function.

\begin{figure}[htb]
\begin{center}
\begin{tabular}{c}
\epsfxsize=10truecm
\epsffile{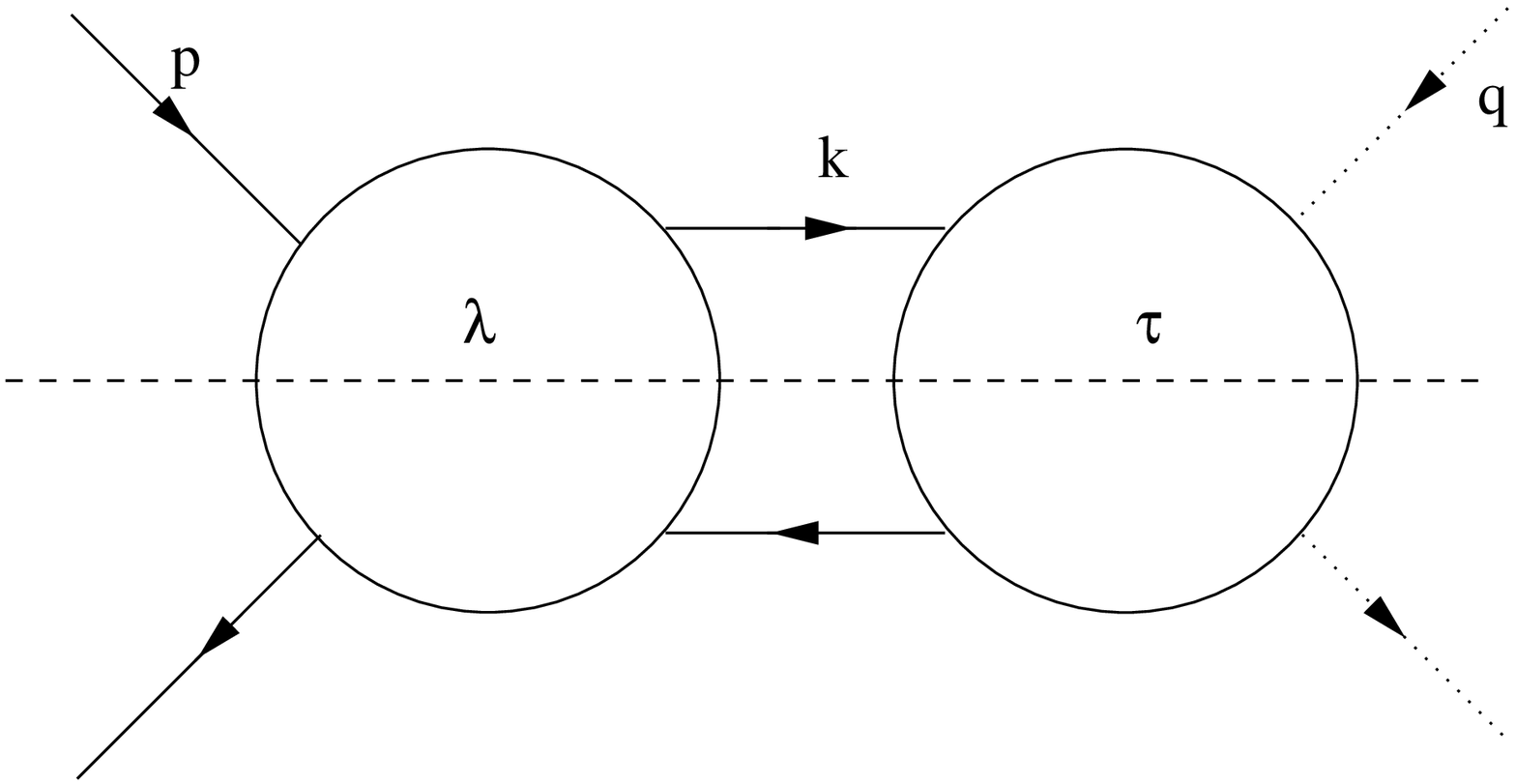}\\
\end{tabular}
\end{center}
\caption{\label{decomposition1}{\em \small Relevant decomposition for DIS in $\pts$}}
\end{figure}

In the following
we will generalize this construction for a general graph:
the leading contributions to the DIS structure function
in $\pts$ are given by the decomposition
depicted in Fig.\ref{decomposition1}.

Here $\tau$ is the hard subgraph, that is the part of the graph in which
a large momentum flows, and $\lambda$ is the soft subgraph.
Decompositions with more than two legs connecting the hard to the
soft subgraph are suppressed by powers of $1/Q^2$. By using this fact and
Zimmermann formalism \cite{zim} we will be able
to define a cut vertex expansion at all orders.

\resub{Cut vertex expansion for DIS in $\pts$}

\begin{figure}[htb]
\begin{center}
\begin{tabular}{c}
\epsfxsize=15truecm
\epsffile{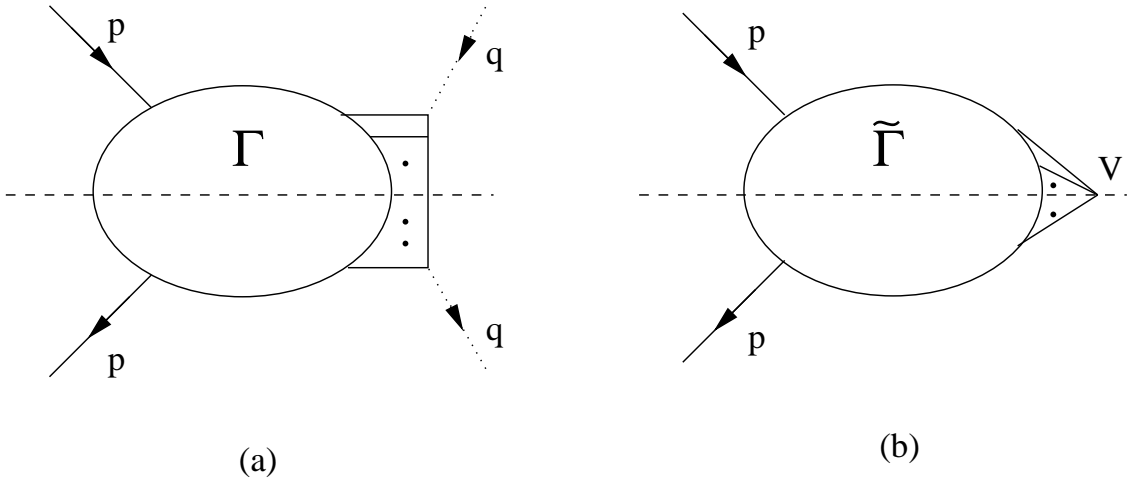}\\
\end{tabular}
\end{center}
\caption{\label{gamma}{\em \small (a) general cut graph $\G$ contributing to the
DIS structure function
(b) the graph obtained from $\G$ by merging the currents at the same
point $V$}}
\end{figure}

Let us consider a general graph $\Gamma$ giving
contribution to the
deep inelastic structure function, renormalized for example
in the MS scheme.
We
saw
that the strategy
which allows
to isolate the leading term in
the structure function
consists in
oversubtractions.
In order to implement this
program
at all orders we need a definition of renormalization
part and of a subtraction operator.
First of all we define the graph ${\tilde \G}$
as 
$\G$ in which
the vertices of insertion
of $J$ are merged
at the same point $V$
(see Fig.\ref{gamma}).

Keeping in mind the conclusions of the previous subsection we define 
a renormalization part $\g$ as a subgraph of
$\G$ such that ${\tilde \gamma}$ is either the bare cut vertex or a proper
subdiagram of ${\tilde \G}$ containing $V$ with two external legs.
As usual a forest $\UU$ of $\Gamma$ is a set of non overlapping renormalization
parts of $\G$. A forest may be empty.

Given a renormalization part $\gamma$ of $\Gamma$, its contribution
in the loop integral
as a function of external momenta must be of
the form $I_\gamma(k,q)$.
We define the subtraction operator $t^\gamma$ as
\begin{equation}
t^\gamma I_\gamma (k,q)= I_\gamma (\kh,q).
\end{equation}
Now,
as we have done in the previous examples, we have
to relate the structure function, the
leading term and the remainder through an identity.

Given the graph $\G$ the contribution to the structure function is
\begin{equation}
W_\G(p,q)=\int  J_\G(p,k_1...k_l,q)
\f{d^6k_1}{(2\pi)^6}...\f{d^6k_l}{(2\pi)^6} \equiv I_\G
\end{equation}
where $k_i,i=1...l$ is the set of loop momenta to be integrated over.
We define
\begin{equation}
X_\G=\sum_{\UU(\G)\neq\emptyset}~\prod_{\g\in \UU} \left(-t^\g\right) I_\Gamma
\end{equation}
\begin{equation}
Z_\Gamma=\sum_{\Ut (\G)}~\prod_{\g\in\Ut} \left(-t^\g\right) I_\Gamma
\end{equation}
\begin{equation}
Y_\Gamma=-\sum_{\Ut (\G)}~\prod_{\g\in\Ut} 
\left(-t^\g\right)\left(-t^{\g_t}\right) I_\Gamma
\end{equation} 
where $\Ut(\G)$ are $\G$ forests which do not contain the bare cut vertex
as renormalization part, and $\g_t$ is the bare cut vertex.
The following identity holds \cite{zim,lev}
\begin{equation}
\label{iden}
W_\G=-X_\G+Z_\G-Y_\G.
\end{equation}
The claim is that
in (\ref{iden}) $-X_\G$ is the leading term while $Z_\G-Y_\G$ is the
remainder and is suppressed in the large $Q$ limit. 

The next step is to show that $X_\G$ has a factorized structure.
The definition
of renormalization part we have given is such that the
renormalization parts in a forest $\UU$ are all nested.
Therefore it is possible to
select the innermost renormalization part $\g_0$ and
the expression for $X_\G$ can be recast
in the form
\begin{equation}
X_\G=-\sum_{\g_0}\sum_{\Ut(\G /\g_0)}~
\prod_{\g\in\Ut(\G /\g_0)} (-t^\g)
\int\dk I_{\G/\g_0}(p,k)~ t^{\g_0} I_{\g_0}(k,q),
\end{equation}
where we
used our definition of renormalization part.
By taking moments we get a completely factorized expression
\begin{align}
\label{mexpansion}
\left(W_\G\right)_\sigma\sim -\left(X_\Gamma\right)_\sigma
&=\sum_{\g_0}\sum_{\Ut(\G /\g_0)}~
\prod_{\g\in\Ut(\G /\g_0)} (-t^\g)
\vt_{\G/\g_0}^\sigma(p^2)~ \Ct^\sigma_{\g_0}(Q^2)\nn\\
&=\sum_{\g_0} \vb_{\G/\g_0}^\sigma(p^2) 
\Cb^\sigma_{\g_0}(Q^2)\equiv v_\sigma^M(p^2) C^M_\sigma(Q^2)
\end{align}

where
\begin{align}
\vt_{\G/\g_0}^\sigma(p^2)&=
\int_0^1 dx x^{\sigma-1}\int I_{\G/\g_0}(p,k)x\delta
\left(x-\f{k_+}{p_+}\right)\dk\nn\\
&=\int I_{\G/\g_0}(p,k)
\left(\f{k_+}{p_+}\right)^{\sigma}\theta(k_+) \dk
\end{align}
and
\begin{equation}
\Ct^\sigma_{\g_0}(Q^2)=\int_0^1 dx x^{\sigma-1} I_{\g_0}(0,x,Q^2)
\end{equation}

are the unrenormalized cut vertex and
coefficient function respectively, while 
$\vb_{\G/\g_0}^\sigma(p^2)$ and $\Cb^\sigma_{\g_0}(Q^2)$
are renormalized quantities\footnote{Actually in
this renormalization scheme
$\Cb^\sigma_{\g_0}(Q^2)=\Ct^\sigma_{\g_0}(Q^2)$}.
We saw in the previous subsection that
in the large $Q$ limit
separately
divergent contributions 
may
arise
even if we
start from a UV finite integral. In eq. (\ref{mexpansion})
the operator
acting on $\vt_{\G/\g_0}^\sigma(p^2)$
performs cut vertex renormalization
by using the same prescription
adopted in
oversubtraction.  
Thus
eq. (\ref{mexpansion}) is
just
the cut vertex expansion proposed
by Mueller in Ref.\cite{mueller}, with the only difference
that in Ref.\cite{mueller} the ordinary UV renormalization
was explicitly performed with zero momentum
subtraction, whereas here we have worked
with renormalized quantities.

In this paper we prefer to
use
minimally
subtracted cut vertices \cite{gupta}, and so
doing
we slightly modify the expression for $X_\G$
by following Ref.\cite{lev}.

From now on we will work in $d=6-2\eps$ dimensions and define the operator
$\tge$ as
acting
on a renormalization part $\g$ by
annihilating all terms which are not singular as $\eps\to 0$.
A particular $\tge$ acts after the loop integrals inside $\g$
are performed,
and
the ordinary UV counterterms
added.

Given a forest $\UU(\G)\neq \emptyset$ call $\tau$ the
outermost $\g \in\UU(\G)$ and let $\UUT(\tau)$ be a forest of $\G$
such that $\g\in \UUT(\tau)\Leftrightarrow \g \supseteq\tau$.
Furthermore, define $\UUT_t(\tau)$ a forest of $\G$
such that $\g\in \UUT_t(\tau)\Leftrightarrow \g \supsetneq \tau$.
The quantity
\begin{equation}
X_\G^\prime=\sum_{\UU(\G)\neq\emptyset}~\sum_{\UUT(\tau)}~
\prod_{\g\in\UUT}(-\tge)\prod_{\g\in\UU}(-t^\g) I_\G
\end{equation}
differs from $X_\G$ in
that it contains extra UV counterterms which
make each forest contribution separately finite. However, $X_\G$
is free from UV divergences, so that extra counterterms in $X_\Gamma^\prime$
must cancel and we have $X^\prime_\G=X_\G$.
$X^\prime_\G$ can be rearranged in the form
\begin{align}
\label{x1}
X_\G^\prime &=-\sum_{\tau}\sum_{\UUT_t(\tau)}
\prod_{\g\in\UUT_t} (-\tge)
(1-t_\eps^{\tau}) t^{\tau}\sum_{\Un(\tau)}\prod_{\g\in\Un} (-t^\g) I_\G\nn\\
&=-\sum_{\tau}\int\dkd\sum_{\UUT_t(\tau)}\prod_{\g\in\UUT_t} (-\tge)
I_{\G/\tau}(p,k)(1-t_\eps^{\tau}) 
t^{\tau}\sum_{\Un(\tau)}\prod_{\g\in\Un} (-t^\g) I_{\tau}(k,q)
\end{align}
where $\Un(\tau)$ are the normal forests of $\tau$, i.e. forests of
$\tau$ which do not contain $\tau$.
Eq.(\ref{x1}) can be easily derived by observing that
\begin{equation}
\prod_{\g\in\UU} (-t^\g) I_\G=(-t^{\tau})\prod_{\g\in\Un(\tau)}
(-t^\g) I_\G
\end{equation}
and that the forest $\UUT (\tau)$ can either contain $\tau$ or not:
\begin{equation}
\sum_{\UUT (\tau)}~\prod_{\g\in\UUT} (-\tge)=\sum_{\UUT_t (\tau)}
~\prod_{\g\in\UUT_t} (-\tge) (1-t_\eps^{\tau}).
\end{equation}
Let us comment on the structure of eq. (\ref{x1}). Given a maximal
renormalization part $\tau$, a decomposition as in
Fig.\ref{decomposition1} follows.
The
operator acting on $I_\tau(k,q)$ performs
subtractions corresponding to possible renormalization parts $\g$
contained into $\tau$.
The operator $t^{\tau}$
replaces $k$ with ${\hat k}$
as in the examples of the previous subsection and $(1-t^{\tau}_\eps)$ removes
possible UV divergences, so that what we get on the right is the coefficient
function. On the left the operator
acting on $I_{\G/\tau}(p,k)$
renormalizes the cut vertex, now in the MS scheme.
Thus,
by taking moments, we can write
\begin{align}
\label{expansion}
\left(W_\G\right)_\sigma\simeq -\left(X_\G\right)_\sigma
&=\sum_{\tau}\sum_{\UUT_t(\tau)}\prod_{\g\in\UUT_t} (-\tge)
\vt^\sigma_{\G/\tau}(p^2) (1-t_\eps^{\tau})
\sum_{\Un(\tau)}\prod_{\g\in\Un} (-t^\g) \Ct_{\tau}^\sigma(Q^2)
\nn\\
&=\sum_{\tau} v^\sigma_{\G/\tau}(p^2)C^\sigma_{\tau}(Q^2)
\equiv v_\sigma(p^2) C_\sigma(Q^2)
\end{align}
that is to say, we end up with a completely factorized expression.
Here $v^\sigma_{\G/\tau}(p^2)$ and $C^\sigma_{\tau}(Q^2)$
are the renormalized cut vertex and coefficient function in the MS scheme respectively.

Finally we want to show how
eq. (\ref{x1}) works with the examples
considered before.

\begin{figure}[htb]
\begin{center}
\begin{tabular}{c}
\epsfxsize=6truecm
\epsffile{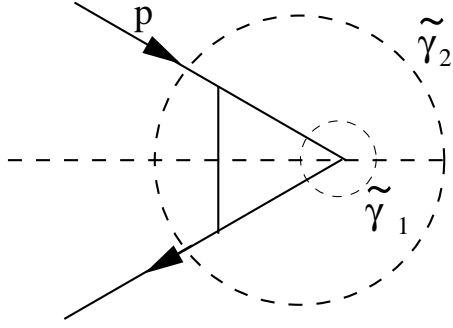}\\
\end{tabular}
\end{center}
\caption{\label{ldr}{\em \small The diagram ${\tilde \G}_1$ and its subdiagrams ${\tilde \g_1}$ and ${\tilde \g_2}$}}
\end{figure}

Let us go back to the diagram in Fig.\ref{lad} and call it $\G_1$.
The diagram ${\tilde \G}_1$ is represented in Fig.\ref{ldr} 
and $\g_1$ and $\g_2$ appear to be the renormalization parts.
From eq. (\ref{x1}) we obtain
\begin{align}
W_1(p,q)\simeq -X_{\Gamma_1}&=\int (1-t_\eps^{\g_2})
I_{\G_1/\g_1}(p,k)t^{\g_1} 
I_{\g_1}(k,q)\dkd\nn\\
&+\int I_{\G_1/\g_2}(p,k)(1-t_\eps^{\g_2})
t^{\g_2} (1-t^{\g_1})I_{\g_2}(k,q)\dkd
\end{align}
By taking moments we get the renormalized
form of eq. (\ref{w1})
in the MS scheme.

\begin{figure}[htb]
\begin{center}
\begin{tabular}{c}
\epsfxsize=6truecm
\epsffile{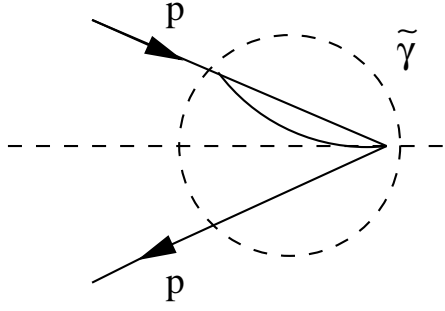}\\
\end{tabular}
\end{center}
\caption{\label{vr}{\em \small The diagram ${\tilde \G}_2$ shows that $\G_2$
has only one renormalization part which is the graph itself}}
\end{figure}

Then consider the diagram in Fig.\ref{virtual} and call it $\G_2$.
The corresponding
diagram ${\tilde \G}_2$ is represented in Fig.\ref{vr}.
The graph itself gives the only renormalization part $\g$
and so eq. (\ref{expansion}) reads
\begin{equation}
W_2(p,q)\simeq\int I_{\G_2/\g}(p,k) t^{\g}I_{\g}(k,q)\dk
=W_2(0,x,Q^2).
\end{equation}

Eventually we should show that the expansion outlined in this section
corresponds indeed to the
leading contribution or, equivalently, the remainder is suppressed by
a power of $1/Q^2$.
Naively
this happens
since all the leading momentum flows
are subtracted off \cite{mueller},
but we will provide a more
precise
argument in Sect. 4.
Nevertheless it can be shown \cite{bau,mun} that
the spacelike cut vertex
represents
the analytic continuation in the spin variable 
of the matrix element of minimal twist operator
$O_{\mu_1...\mu_n}=\p (0)\d_{\mu_1}...\d_{\mu_n}\p (0)$.
Thus the
cut vertex expansion is equivalent in this case to OPE
and we do not need to
exhibit
any additional proof.

\resection{Generalized cut vertex expansion}

\resub{Examples}

Here we are going to
discuss the process $p+J(q)\to p^\prime+X$.
In this case a semi-inclusive structure function can be defined
\begin{equation}
W(p,p^\prime,q)=\f{Q^2}{2\pi} \sum_X\int d^6x e^{iqx} <p|J(x)|p^\prime X>
<X p^\prime|J(0)|p>
\end{equation}
and let $z$ be
\begin{equation}
z=\f{p^\prime q}{pq}.
\end{equation}

\begin{figure}[htb]
\begin{center}
\begin{tabular}{c}
\epsfxsize=5truecm
\epsffile{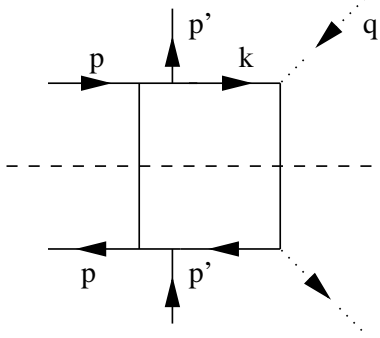}\\
\end{tabular}
\end{center}
\caption{\label{sdis1}{\em \small A one loop contribution to the semi-inclusive structure function}}
\end{figure}

In Ref.\cite{cut} it was argued that in the region $t\ll Q^2$ an expansion similar
to (\ref{expansion}) holds for the semi-inclusive structure function.
In Ref.\cite{graz} a one loop calculation of the semi-inclusive cross section
has been performed which confirms such expectation. We will now point out
how to perform such a generalization,
and, as we did before, we will start with some examples.
Consider the diagram in Fig.\ref{sdis1},
which takes into account real emission from the initial leg.

We have
\begin{equation}
W_1(p,p^\prime,q)=\int V_1(p,p^\prime,k) M_1(k,q) \dk
\end{equation}
where
\begin{equation}
V_1(p,p^\prime,k)=\lambda^42\pi\delta_+\left((p-k-p^\prime)^2\right)\f{1}{k^4}
\f{1}{(k+p^\prime)^4}
\end{equation}
and $M_1(k,q)$ is given by eq. (\ref{m1}).
In the region $t\ll Q^2$
the integral gets its leading contribution in the region of small transverse
momentum ${\bf k}_\perp$ \cite{graz}, and so we can write
\begin{equation}
W_1(p,p^\prime,q)\simeq \int V_1(p,p^\prime,k) M_1({\hat k},q)\dk.
\end{equation}
If we define
\begin{equation}
v(p,p^\prime,\xb)=\int V_1(p,p^\prime,k)\xb\delta
\left(\xb-\f{k_+}{p_+-p^\prime_+}\right)\dk
\end{equation}
where
\begin{equation}
{\bar x}=\f{x}{1-z}
\end{equation}
we easily find
\begin{equation}
W(p,p^\prime,q)=\int \f{du}{u} v(p,p^\prime,u)C_1(\xb/u,Q^2)
\end{equation}
where $M_1({\hat p},q)\equiv M_1(0,x,Q^2)=C_1(x,Q^2)$
is the same lowest order coefficient function appearing in
eq.(\ref{w1}).
 
\begin{figure}[htb]
\begin{center}
\begin{tabular}{c}
\epsfxsize=6truecm
\epsffile{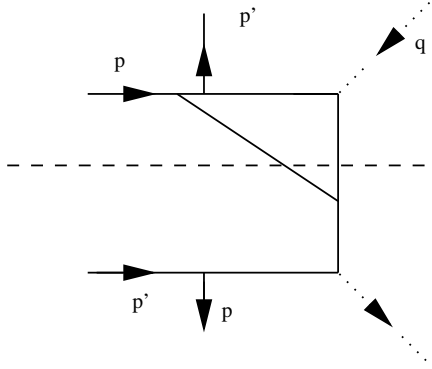}\\
\end{tabular}
\end{center}
\caption{\label{int}{\em \small A contribution to
the semi-inclusive structure function which represents interference between current and target fragmentation}}
\end{figure}
 
Let us consider the diagram in Fig.\ref{int},
representing an interference
between real emission from the initial and final
legs. It turns out that this graph is suppressed by a power of
$t/Q^2$ in the limit $t\ll Q^2$ and so can be neglected.
This result stems from the fact that in the large $Q^2$
limit three lines go into the hard
subgraph, which produces a factor $t/Q^2$.

These simple examples, together with the calculation performed in Ref.
\cite{graz}, suggest that a cut vertex
expansion can be constructed in this case as a simple
generalization of the formalism previously outlined.

\resub{General case}

\begin{figure}[htb]
\begin{center}
\begin{tabular}{c}
\epsfxsize=15truecm
\epsffile{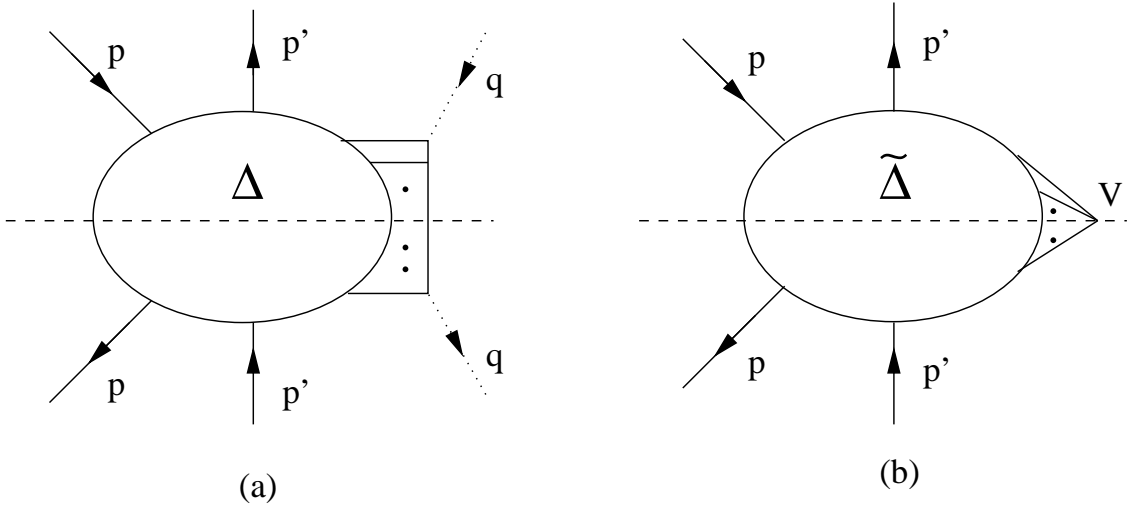}\\
\end{tabular}
\end{center}
\caption{\label{delta}{\em \small (a) General cut graph $\D$ for semi-inclusive DIS in
$\pts$ (b) the graph obtained from $\D$ by merging the currents at the
same point $V$ }}
\end{figure}

\begin{figure}[htb]
\begin{center}
\begin{tabular}{c}
\epsfxsize=10truecm
\epsffile{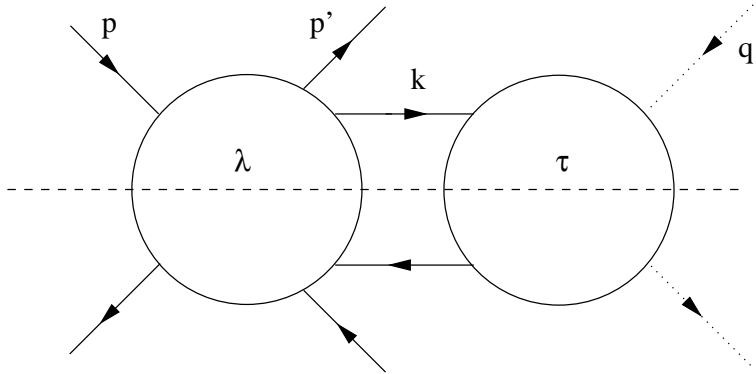}\\
\end{tabular}
\end{center}
\caption{\label{decomposition2}
{\em \small Relevant decomposition for the semi-inclusive structure
function in the region $t\ll Q^2$}}
\end{figure}

The general graph $\D$ which contributes to the semi-inclusive structure
function is depicted in Fig.\ref{delta}(a).
In the kinematic region under
study we can build up a cut vertex expansion via a
straightforward
generalization of the formalism given in the previous section. We again
suppose that
the ordinary UV divergences have been removed with suitable
counterterms. As we did
in the inclusive case
we define renormalization parts looking at the
diagram ${\tilde \D}$ obtained
by merging the two external currents at the same
point $V$ (see Fig.\ref{delta}(b)). Therefore
a renormalization part $\g$ is again a subgraph of $\D$ such that
${\tilde \g}$ is either the bare cut vertex or a proper subdiagram
of ${\tilde \D}$ with two external legs.
An identity analogous to
eq. (\ref{iden}) holds in this case and the expansion has
exactly the same structure. What we claim is that
in the MS scheme the leading term reads
\begin{equation}
\label{sx1}
X_\D^\prime=-\sum_{\tau}\sum_{\UUT_t(\tau)}\prod_{\g\in\UUT_t} (-\tge)
(1-t_\eps^{\tau}) t^{\tau}\sum_{\Un(\tau)}\prod_{\g\in\Un} (-t^\g) I_\D
\end{equation}
or, by taking moments, now with respect to $\xb$,
\begin{align}
\label{sexpansion}
\left(W_\Delta\right)_\sigma &\simeq\sum_{\tau}\sum_{\UUT_t(\tau)}\prod_{\g\in\UUT_t} (-\tge)
\vt^\sigma_{\D/\tau}(p,p^\prime) (1-t_\eps^{\tau})
\sum_{\Un(\tau)}\prod_{\g\in\Un} (-t^\g) 
\Ct_{\tau}^\sigma(Q^2)\nn\\
&=\sum_{\tau} v^\sigma_{\Delta/\tau}(p,p^\prime) 
C_\sigma^{\tau} (Q^2)\equiv
v_\sigma(p,p^\prime)
C_\sigma(Q^2)
\end{align}
which corresponds to consider only the
decomposition in Fig.\ref{decomposition2}.
\vspace*{1truecm}

Here $v_\sigma(p,p^\prime)$ is a
generalized cut vertex \cite{cut}, with four external legs,
renormalized in the MS scheme and
$C_\sigma(Q^2)$ is its coefficient function.
As pointed out in Ref.\cite{cut},
the coefficient function coincides with that of
the inclusive case,
since they both originate from the hard part of the graphs.

By saying that (\ref{sexpansion}) is the
leading term in the semi-inclusive cross section we mean that
corrections are suppressed by powers of $p^2/Q^2$, $p^{\prime 2}/Q^2$,
$t/Q^2$.
\begin{figure}[htb]
\begin{center}
\begin{tabular}{c}
\epsfxsize=6truecm
\epsffile{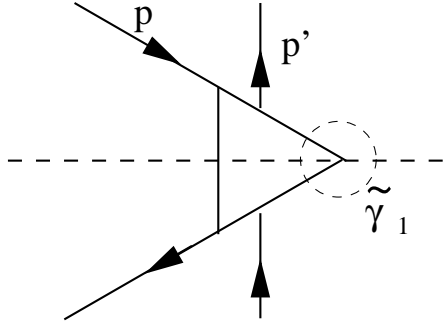}\\
\end{tabular}
\end{center}
\caption{\label{sdr}{\em \small The diagram ${\tilde \D}_1$ and its renormalization
part ${\tilde \g_1}$}}
\end{figure}

Now, following the previous section,
we will show some applications of eq. (\ref{sx1}).
Let us go back to the graph in Fig.\ref{sdis1} and call it $\D_1$.
The graph
${\tilde \D}_1$ is depicted in Fig.\ref{sdr} and $\g_1$ appears as the only
renormalization part. Therefore from eq. (\ref{sx1}) we get
\begin{align}
W_1(p,p^\prime,q)&\equiv\int V_1(p,p^\prime,k) M_1(k,q)\dkd\nn\\
&\simeq \int I_{\D/\g_1}(p,p^\prime,k) t^{\g_1} I_{\g_1}(k,q)\dkd
=\int V_1(p,p^\prime,k)M_1(\kh,q)\dkd\nn\\
&=\int \f{du}{u} v(p,p^\prime,u)C(\xb/u,Q^2).
\end{align}

Consider now the graph in Fig.\ref{int} and call it $\D_2$. 
This graph does not have renormalization parts, so
(\ref{sx1}) gives
\begin{equation}
W_2(p,p^\prime,q)\simeq -X_{\D_2}=0
\end{equation}
at
leading power.

Up to now we have considered only contributions of target fragmentation
or at least interference between target and current fragmentation.
\begin{figure}[htb]
\begin{center}
\begin{tabular}{c}
\epsfxsize=10truecm
\epsffile{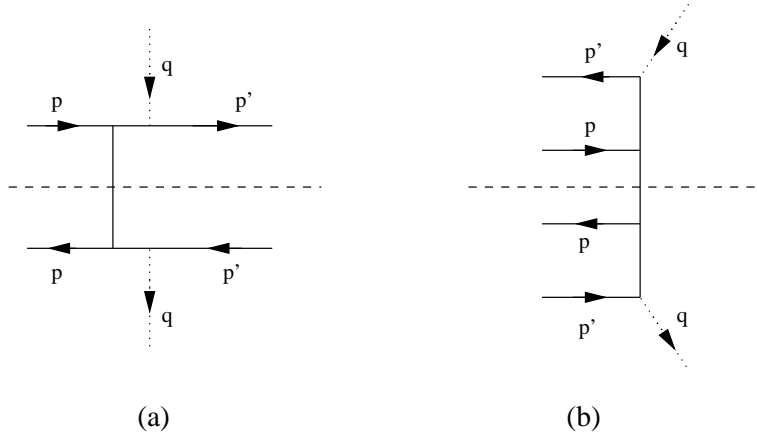}\\
\end{tabular}
\end{center}
\caption{\label{current}{\em \small (a) A current fragmentation contribution to the semi-inclusive structure function (b) The same graph depicted in a different form}}
\end{figure}

One may wonder how to treat a graph as
in Fig.\ref{current}(a), say $\D_3$.
Here the observed
particle comes from current fragmentation. Nevertheless this
graph can be recast in
the form of Fig.\ref{current}(b) and appears to have no renormalization
parts. Hence eq. (\ref{sx1}) gives again
\begin{equation}
W_{\D_3}(p,p^\prime,q)\simeq -X_{\D_3}=0
\end{equation}
still at leading power.

We have seen so far how the expansion works also in the
semi-inclusive case.
We have now to prove that $-X_\D$ is
the leading contribution, while $W_\D-(-X_\D)$ is
suppressed by powers of $1/Q^2$. Notice that in this case
we cannot use
OPE as in inclusive DIS to justify the statement.

\resection{Factorization}

In this section we prove in general that
the expansion we
have
formulated for the process
$p+J(q)\to p^\prime+X$
in the limit $t \ll Q^2$
gives
the leading
term in the cross section.
The argument we will
provide
works also for inclusive DIS,
though in that case OPE can be
safely
used.

We will strongly rely on the ideas of Ref.\cite{css}.
Let us consider the semi-inclusive structure function and
suppose we have scaled it of an overall factor in order to make
it dimensionless. We can write
\begin{equation}
W=W\left(x,z,p^2/\mu^2,p^{\prime 2}/\mu^2,t/\mu^2,Q^2/\mu^2,
\alpha(\mu^2)\right).
\end{equation}
Since
the renormalization scale is arbitrary,
we can get insight into
the high $Q^2$ limit
by setting $\mu=Q$
and looking at the singularities in the limit $p^2$, $p^{\prime 2}$,
$t\to 0$. The singularities of a general Feynman graph in this
limit arise only from
definite
regions of integration in loop
momenta, around surfaces called {\em pinch surfaces}
and can be found
through
Landau equations \cite{Landau}.
From a general Feynman diagram at a particular pinch surface
one can define a {\em reduced} diagram in which:

\begin{itemize}
\item all far off shell
lines are contracted to a point 
\item all collinear lines in
a given direction are grouped into subdiagrams called {\em jets}
\item soft lines ($k\ll Q$) are grouped into a soft diagram which is
arbitrarily
connected to the rest of the graph.  
\end{itemize}

\begin{figure}[htb]
\begin{center}
\begin{tabular}{c}
\epsfxsize=8truecm
\epsffile{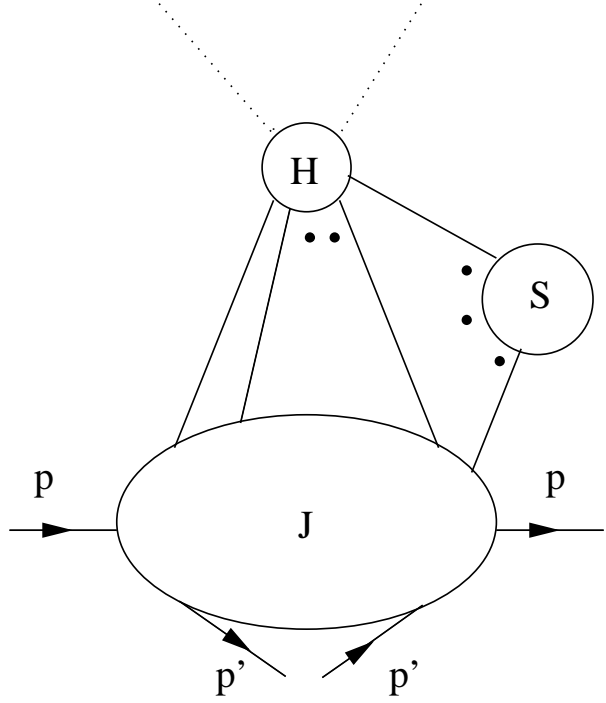}\\
\end{tabular}
\end{center}
\caption{\label{reduced}{\em \small General reduced graph for semi-inclusive DIS in $\pts$}}
\end{figure}

The general reduced graphs for the semi-inclusive process
allowed
by the criterion of physical propagation \cite{cn}
are shown in Fig.\ref{reduced}. 
They involve a jet $J$ in the direction of the incoming particle, from which
the final collinear particle emerges,
a hard part $H$ in
which momenta of order $Q$ circulate and
a soft subgraph $S$.
The jet $J$ is
connected
to $H$ by an arbitrary number of lines $N_{JH}$ and the soft
subdiagram $S$ is
connected both
to $H$ and to $J$,
by $N_{HS}$ and $N_{JS}$ soft lines respectively.
The behavior of the reduced graph $R$ near the pinch surface can be
estimated through infrared power counting \cite{sterman}.
After defining an integration variable $\lambda$ which gives the scaling of
normal variables near the pinch surface the reduced diagram is
${\cal O}\left(\lambda^{\omega(R)}\right)$ where
\begin{equation}
\omega(R)=\f{1}{2} D\hs L_C-N_C+D\hs L_S-2N_S+N_2.
\end{equation}
Here $L_C$ ($L_S$) is the
number of
collinear (soft) loops, $N_C$ ($N_S$) is the number
of collinear (soft) lines, and $N_2$ is the
number of two-point subdiagrams in $R$. Leading behavior occurs
when $\omega(R)$ is minimum.
Furthermore let $L_{C0}$ ($L_{S0}$)
be the number of loops of $J$ ($S$),
$N_{C0}$ ($N_{S0}$) the number of internal lines of $J$ ($S$)
and finally $N_E$ the number of external lines of
$J$.

For the sake of simplicity
we start from the case in which there are no soft lines.
We have ($D=6$)
\begin{align}
\label{omegar}
\omega(R)&=3L_C-N_C+N_2=3(L_{C0}+N_{JH}-1)-N_C+N_2=\nn\\
&=3(N_{C0}-V_J+1+N_{JH}-1)-N_{C0}-N_{JH}+N_2\nn\\
&=2N_{C0}-3V_J+2N_{JH}+N_2\nn\\
&=2N_{C0}-(2N_{C0}+N_{JH}+N_E)+2N_{JH}+N_2\nn\\
&=N_{JH}-N_E+N_2
\end{align}
where we made use of Eulero identity, holding for a general graph
\begin{equation}
\mbox{Loops}=\mbox{Lines}-\mbox{Vertices}+1
\end{equation}
and of the general identity
\begin{equation}
2n+N=3V
\end{equation}
in which $n$ and $N$ are the number of internal
and external lines respectively, $V$ is the number
of vertices and $3$ is the number of
lines going into a vertex. The last identity tells that a line
starts and ends up into a vertex.

It appears from eq. (\ref{omegar}) that for $N_E=2$ (DIS)
we have leading behavior
only with $N_{JH}=2$, or, equivalently, we get leading regions only when two
collinear lines connect the hard to the jet subgraph.
In this case $\omega(R)=0$, i.e. the singularity
is logarithmic. In the case
$N_E=4$ (semi-inclusive DIS) we again get leading behavior when
$N_{JH}=2$ and in this case
$\omega(R)=-2$ , so that the singularity is
power-like\footnote{This result
agrees with the one loop calculation
performed in Ref.\cite{graz} where the
leading contribution to the
semi-inclusive structure function was found to be ${\cal O}(1/t^2)$}.

Now we show that the possibility of a soft graph is ruled
out by power counting.
If we add a soft diagram as in Fig.\ref{reduced} we have to add to
eq. (\ref{omegar}) the term
\begin{align}
6L_S-2N_S&=6(L_{S0}+N_{JS}+N_{HS}-1)-2(N_{S0}+N_{JS}+N_{HS})\nn\\
&=6(N_{S0}-V_S+1+N_{JS}+N_{HS}-1)-2(N_{S0}+N_{JS}+N_{HS})\nn\\
&=4N_{S0}-6V_S+4(N_{JS}+N_{HS})\nn\\
&=4N_{S0}-(4N_{S0}+2N_{JS}+2N_{HS})+4(N_{JS}+N_{HS})\nn\\
&=2(N_{JS}+N_{HS})
\end{align}
where we used
the same identities as before. Hence we get
\begin{equation}
\omega(R)=N_{JH}-N_E+2(N_{JS}+N_{HS})+N_2
\end{equation}
and the absence of soft lines in a leading graph is manifest.

\begin{figure}[htb]
\begin{center}
\begin{tabular}{c}
\epsfxsize=8truecm
\epsffile{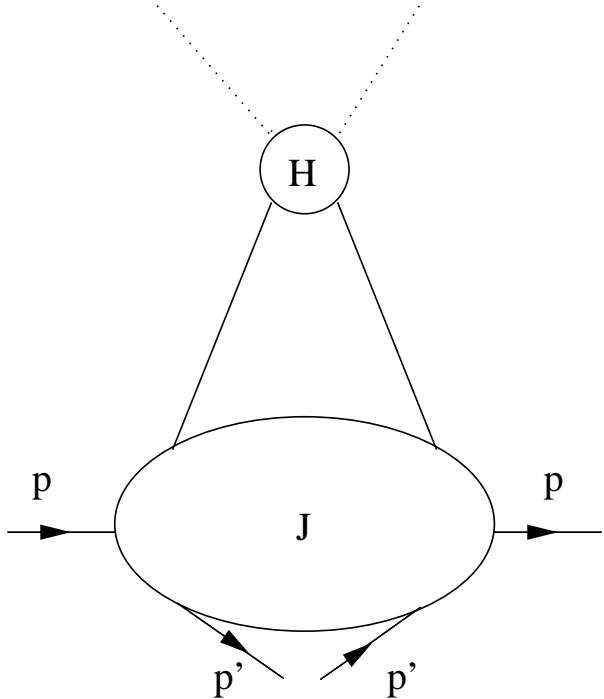}\\
\end{tabular}
\end{center}
\caption{\label{lead}{\em \small Leading regions for semi-inclusive DIS in $\pts$}}
\end{figure}

It follows that the leading regions for semi-inclusive deep inelastic
scattering are of the form of Fig.\ref{lead} \cite{cut}.
Thus, being this statement exactly
equivalent
to say that the relevant decomposition is that of
Fig.\ref{decomposition2}, we can conclude that our generalized cut vertex expansion really
gives the leading contribution to the cross section.

\resection{Conclusions}

In this paper we
constructed
a generalization of the cut vertices
formalism in order to deal with semi-inclusive
deep inelastic
processes in the target fragmentation region. We
showed that this program can be achieved without substantial
complications. We can write down an identity which involves the
structure function, the leading term and the remainder. The first
step is to show that the leading term has a factorized structure.
The second step is to
prove
that the leading term is {\em really} leading,
i.e., that the remainder is suppressed by a power of $1/Q^2$.
With this purpose in mind
we
followed the ideas of
Ref.\cite{css}, showing that the leading regions for the
semi-inclusive cross section in the limit $t\ll Q^2$ are
of the same form as in inclusive DIS.

The formalism proposed in Ref.\cite{mueller}
has been slightly modified so as
to obtain
an expansion
in terms of minimally subtracted cut vertices. We think this
necessary in order to get well-behaved coefficient functions in the
zero mass limit.

Although the model in which we have been working
is $\pts$ and not QCD, the scalar $\pts$ model
is an excellent framework to discuss
the properties of strong interactions
at short distances
as it resembles QCD in a
physical gauge. However in QCD
further complications arise due to soft gluon lines
connecting the hard to the
jet subdiagrams, which are not suppressed as
in $\pts$ by power counting.
One has to use Ward identities to show that
they cancel out. Such an issue  was beyond the
aim of this paper
and we expect that this complication
will not spoil factorization
as was also argued in Refs.\cite{cut,soper}.

{\em Note added:} After the completion of this work,
a paper by Collins appeared \cite{collins} in which
the proof of factorization for diffractive hard scattering
in QCD is presented. The proof is based on the cancellation
of soft gluons exchanges between the hard and
the jet subdiagram, and applies as well to the process we
have been considering here. The results obtained by Collins
completely agree with ours and
confirm the validity of our approach.

\vspace{3mm}
\begin{center}
\begin{large}
{\bf Acknowledgments}
\end{large} 
\end{center}
\noindent I am indebted to
S. Catani, J.C. Collins, J. Kodaira,
G. Oderda, G. Sterman, G. Veneziano and particularly to D.E. Soper for
helpful discussions, and to F. Vian for carefully reading
the manuscript. Special thanks
to L. Trentadue: without
his continuous encouragement this work wouldn't have been possible.

\end{document}